# Mapping of Ebola virus spillover: suitability and seasonal variability at the landscape scale

Larisa Lee-Cruz [1,2,3,4], Maxime Lenormand [4], Julien Cappelle [1,2], Alexandre Caron [1,2,5], Hélène De Nys [2,6], Martine Peeters [7], Mathieu Bourgarel [2,6], François Roger [1,2], Annelise Tran [1,2,3,4] *

[1] CIRAD, UMR ASTRE, Montpellier, France

[2] ASTRE, Univ Montpellier, CIRAD, INRAE, Montpellier, France

[3] CIRAD, UMR TETIS, Montpellier, France

[4] TETIS, Univ Montpellier, AgroParisTech, CIRAD, CNRS, INRAE, Montpellier, France

[5] Faculdade Veterinaria, Universidade Eduardo Mondlane, Maputo, Mozambique

[6] CIRAD, UMR ASTRE, Harare, Zimbabwe

[7] TransVIHMI, IRD, INSERM, Univ Montpellier, Montpellier, France

* annelise.tran@cirad.fr

## Abstract

The unexpected Ebola virus outbreak in West Africa in 2014 involving the *Zaire ebolavirus* made clear that other regions outside Central Africa, its previously documented niche, were at risk of future epidemics. The complex transmission cycle and a lack of epidemiological data make mapping areas at risk of the disease challenging. We used a Geographic Information System-based multicriteria evaluation (GIS-MCE), a knowledge-based approach, to identify areas suitable for Ebola virus spillover to humans in regions of Guinea, Congo and Gabon where Ebola viruses already emerged. We identified environmental, climatic and anthropogenic risk factors and potential hosts from a literature review. Geographical data layers, representing risk factors, were combined to produce suitability maps of Ebola virus spillover at the landscape scale. Our maps show high spatial and temporal variability in the suitability for Ebola virus spillover at a fine regional scale. Reported spillover events fell in areas of intermediate to high suitability in our maps, and a sensitivity analysis showed that the maps produced were robust. There are still important gaps in our knowledge about what factors are associated with the risk of Ebola virus spillover. As more information becomes available, maps produced using the GIS-MCE approach can be easily updated to improve surveillance and the prevention of future outbreaks.

## Author Summary

Short Title: Risk mapping of Ebola virus spillover

Ebola virus disease is a highly pathogenic disease transmitted from wildlife to humans. It was first described in 1976 and its distribution remained restricted to Central Africa until 2014, when an outbreak in West Africa, causing more than 28,000 cases and more than 11,000 deaths, took place. Anthropogenic factors, such as bushmeat hunting, trade and consumption, and environmental and climatic factors, may promote the contact between humans and infected animals, such as bats, primates and duikers, increasing the risk of virus transmission to the human population. In this study, we used the spatial multicriteria evaluation framework to gather all available information on risk factors and animal species susceptible to infection, and produce maps of areas suitable for Ebola virus spillover in regions in Guinea, Congo and Gabon. The resulting maps highlighted high spatial and temporal variability in the suitability for Ebola virus spillover. Data from reported cases of Ebola virus transmission from wild animals to humans were used to validate the maps. The approach developed is capable of integrating a wide diversity of risk factors, and provides a flexible and simple tool for surveillance, which can be updated as more data and knowledge on risk factors become available.

## Introduction

Ebola virus disease is an emerging zoonotic disease caused by a filovirus with a mortality rate in humans that can reach 90% [1]. There are six species of *Ebolavirus*: *Zaire ebolavirus*, *Sudan ebolavirus*, *Bundibugyo ebolavirus*, *Taï ebolavirus*, *Reston ebolavirus* and *Bombali ebolavirus*. Of these, the first four can infect humans, although only one case of infection by *Taï ebolavirus* has been reported [2]. *Zaire ebolavirus* (EBOV) is apparently the most pathogenic for humans and has been responsible for the majority of outbreaks [1]. The first outbreak of Ebola virus disease took place in 1976 in what is now the Democratic Republic of Congo (then Zaire). There have been 29 outbreaks since, most located in Central Africa, mainly in the Congo basin [3] where the second largest Ebola outbreak, with over 3,000 infections, was declared over in June 2020. A new outbreak was reported in Guinea in January 2021 [4].

The first EBOV outbreak in West Africa took place from 2014-2016. During this outbreak, over 28,000 persons became infected and more than 11,000 died [3]. Recent studies have shown that antibodies against EBOV were present in one person out of 1,483 sampled in 2012 in Guinea [5]. However, genetic studies suggest that the outbreak resulted from a single transmission from the natural reservoir and that the new strain of EBOV was probably introduced from Central Africa during the decade prior to the outbreak [6]. The unexpected onset of the outbreak, thousands of kilometers from previous outbreak locations in Central Africa, and its gravity have made evident the need to better understand which factors might be associated with the spillover of the virus from wildlife to human populations and to identify areas at risk of future outbreaks.

Emerging infectious diseases are diseases caused by the spillover of a pathogen to a new host species population. The pathogen must be able to enter the cells of the new host, replicate and infect more cells within this host [7]. For an epidemic to occur, the onward transmission of the virus between individuals of the new host population must take place (i.e., the virus must be shed and be able to infect more individuals within the novel host population) [8]. Zoonoses are diseases in which the pathogen is transmitted from a vertebrate animal to a human. In several zoonotic infectious diseases including Ebola, the virus can 'jump' from the maintenance (or reservoir) population to multiple host populations including a human population, defined here as the target population (i.e., the population of interest, as defined by Haydon and collaborators [9]). The spillover of a multi-host virus is complex and involves multiple factors and mechanisms. First, the maintenance and target individuals must come into contact, directly (e.g., physical) or indirectly (e.g., through the environment or another intermediate/bridge host). The level of virus circulating and shedding in the maintenance population then must be sufficient at the time of contact to infect the target host. Finally, the target host must be susceptible to infection at the time of contact [10]. In the case of the spillover of *Ebolavirus*, the maintenance species is not known with certainty. However, bats are strongly suspected [11-15], and other wild animals such as duikers and primates have been implicated as intermediate (or bridge) hosts in spillover events to humans [16,17]. Climatic, environmental and anthropic factors might promote contacts between wildlife and humans at different moments in time and space, increasing the probability of a spillover to occur.

Several studies have tried to identify which factors are associated with the transmission of the virus from the maintenance to the intermediate or target hosts, and have mapped areas at risk for *Ebolavirus* spillover. For example Pigott et al. [18,19] mapped the environmental suitability for *Ebolavirus* transmission in Africa based on fruit bats predicted distribution and environmental variables, of which vegetation, temperature and elevation were the most significant. Schmidt et al. [20] included human population density, vegetation and rainfall as factors, and their model showed certain seasonality in the risk of *Ebolavirus* spillover across regions of sub-Saharan Africa. Other studies have pointed to forest loss and fragmentation, which might increase contact between wildlife and humans, and thus favor spillover events [21,22], or to climatic variables that might affect species distribution or the reservoir-virus dynamics [16,23]. All of these studies modelled favorable areas of *Ebolavirus* spillover and/or occurrence at the continental or national scale. However, such large geographic scales might mask heterogeneity in the risk factors and their interactions [24]. For example, factors that may promote contact between the maintenance host and the human population, such as animal and human movements, may appear uniform at large spatial scales, whereas in reality they can be variable at finer scales. Moreover, the relationship between the risk of disease transmission and risk factors may be scale-dependent, as has been suggested for Lyme disease and biodiversity [25]. Therefore, methods applied at a finer spatial scale might provide insight into which local factors are more relevant for the emergence of a virus.

Here we use a Geographic Information System-based multicriteria evaluation (GIS-MCE) to create maps of suitability for EBOV spillover to the human population at a landscape scale. This method has proved useful in epidemiological risk mapping when a lack of epidemiological data exists for a particular disease (e.g., [26,27]). It is based on assumed or existing knowledge of the association between risk factors and the outcome, extracted from expert consultation (e.g., [28,29]) or from a literature review (e.g., [26,30]). First, risk factors are identified and geographical data representing each of the factors are collected. Next, factors are standardized on a continuous scale by transformations or by functions that describe the known or hypothesized relationship between the risk factor and the outcome (i.e., suitable areas for spillover of EBOV in our study). Then, a weight is attributed to each factor before combining them to create a final suitability map.

In this study, we focused on the region of Guinée forestière, in Guinea, where the spillover that caused the 2014-2016 outbreak in West Africa is assumed to have taken place, and two other regions, one in Gabon and one in the Republic of Congo, where previous spillover events have been reported. For our study site in Guinea, we created suitability maps for EBOV at different times to assess the temporal variability in suitable areas. Datasets on the location of reported EBOV spillover events in the three sites were used to validate the suitability maps of EBOV spillover. We also created suitability maps of *Ebolavirus* maintenance in bats as an additional way to assess the performance of the method.

## Methods

### Study area

Our main study area was the forested region in Guinea (i.e., Guinée forestière) for mapping suitable areas for EBOV spillover (Fig 1). This is one of the four natural regions of Guinea, located in the southeast (bordering Sierra Leone and Liberia). We also chose an area in Congo covering part of the Ouest-Cuvette and Cuvette provinces, and an area in Gabon encompassing part of the Woleu-Ntem and Ogooué-Ividno provinces, for validation of our map of suitability for EBOV spillover (see *Map Validation* section).

Indeed, only one EBOV spillover case has been documented in Guinée forestière, in December 2013, while at least 32 cases of spillover to animals and humans have been reported in the combined study areas of Congo and Gabon [16,20,31,32]. The source of the 2021 outbreak in Guinée forestière has not been identified yet [4], although it is more likely a resurgence from a persistently infected survivor than a new spillover event [33]. The three areas are similar in size (Guinée forestière: 42,736 km$^2$; area in Congo: 42,510 km$^2$; area in Gabon: 46,815 km$^2$; Fig 1).

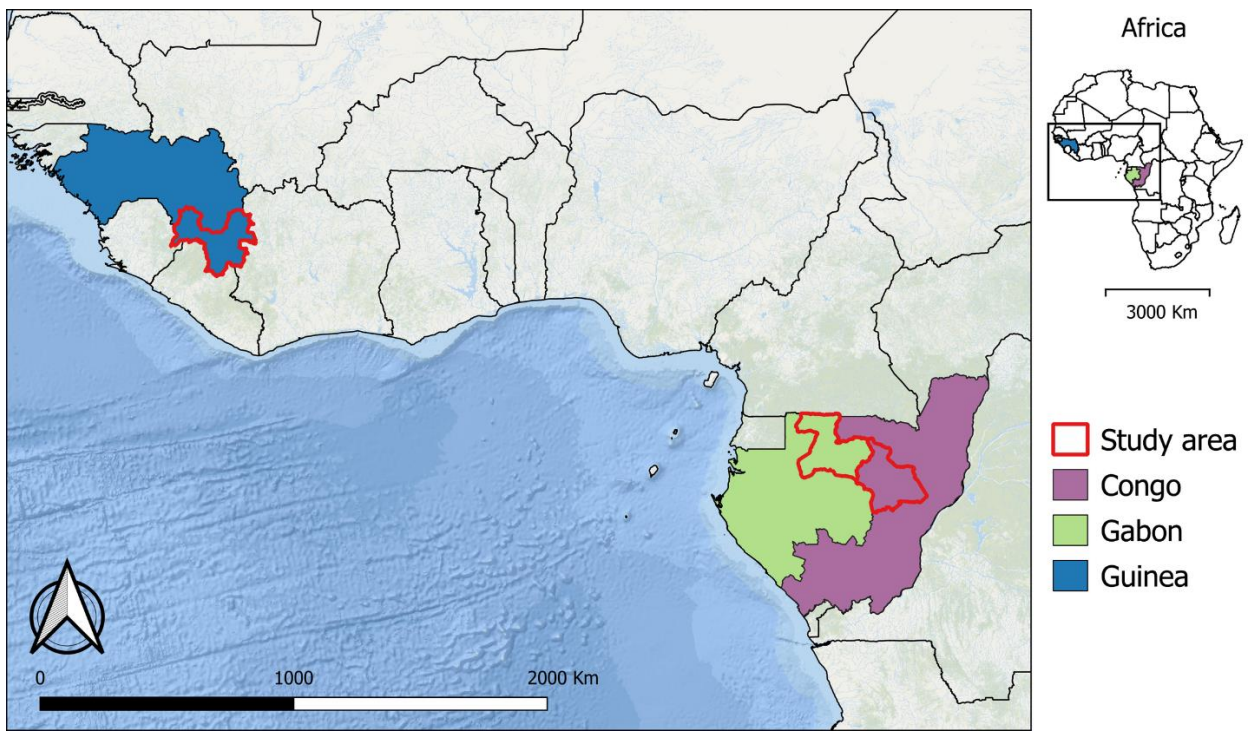

**Fig 1.** Study areas in Guinea, Congo and Gabon. Background: Natural Earth / with Shaded Relief and Water (https://www.naturalearthdata.com)

Guinée forestière has an equatorial humid climate, with strong seasonal precipitation (annual rainfall of over 1,700 mm [34]). Temperatures vary between 30°C and 33°C during the rainy season and between 21°C and 24°C in the dry season [35]. The rainy season is longer than in the rest of the country and can start as early as February and end as late as November. Forests in the region are mainly tropical evergreen forests, but dense forests have been largely substituted by forest-agricultural mosaics [36]. The areas in Gabon and Congo are part of the Congo basin; they are mostly covered by tropical primary and secondary rainforest and semi-evergreen forests. The climate is equatorial with two rainy seasons, March to June and September to December. Temperature ranges from 22°C to 30°C, with an average annual rainfall of 1,500 mm [37,38]. Population density in the chosen areas is low, just over 40 inhabitants/km$^2$ in Guinée forestière, and less than five inhabitants/km$^2$ in the areas in Congo and Gabon.

**Selection of risk factors**

We selected risk factors for EBOV spillover based on a literature review. We mainly focused on EBOV as it is the most pathogenic for humans with 70 % to 90 % mortality rates [1]; however, some of the studies reviewed did not distinguish virus species among infectious *Ebolavirus*. We used the Web of Science to search for studies published between 1990 and 2019, using "ebola", "ecology*" AND "spillover" as keywords. This returned 22 research papers. To increase the number of studies, we conducted a supplementary search in Google Scholar with the same keywords plus "risk" and "zaire" for the same period. This returned 640 results of which we looked into the first 150, as later studies were less relevant for our research question. The selection of Google Scholar was based on a comparative search in different databases in which it performed better, despite bias such as the influence of user's location and previous searches (S1 Table). From these studies, we identified four categories of factors important for the risk of EBOV spillover: 1) the presence of potential animal reservoirs or intermediate hosts implicated in previous spillover events, 2) environmental factors, 3) climatic factors, and 4) factors associated with the trade or consumption of bushmeat. Factors identified in each category are shown in Table 1.

**Formatting and standardization of spatial data**

For each risk factor, we identified an associated variable for which a geographical data layer could be collected (S1 Text). Only open access geographical data were used for ease of reproducibility. The sources of data used are shown in Table A in S1 Text. The identified risk factors, their associated variable and the resolution of the data collected are shown in Table B in S1 Text. Some of the risk factors were considered invariable in the period that we studied, namely the distribution of species, land cover factors (i.e., forest cover, croplands, rivers, roads) and those associated with bushmeat consumption (i.e., hunting areas and villages). Forest cover loss was an estimate of loss between 2001 and 2012 (see Table C in S1 Text). Annual estimates of human population, mean temperature and temperature range, and monthly estimates of Normalized Difference Vegetation Index (NDVI) and rainfall were used.

To be combined in the GIS-MCE, the collected geographical data were transformed to produce standardized spatial risk factors in the same format (raster) and with the same spatial resolution. Spatial data manipulations are detailed in S1 Text and Table C in S1 Text. In the end, each risk factor was represented by a raster image of 1 km x 1 km spatial resolution scaled from 0 (completely unsuitable) to 1 (completely suitable). Maps for each risk factor for the study areas in Guinea, Congo and Gabon are shown in Figs B-D in S1 Text.

**Table 1.** Factors associated to the risk of *Ebolavirus* spillover used in the GIS-MCE. Similar factors (in parenthesis) were considered as one factor for the classification. Some studies did not provide a significance value for the factors (NA). Significance of the risk factor is shown as NS: non-significant /*: p < 0.05 or ≤ 33% of variation explained /**: p < 0.01 or 33% to 66% of variation / ***: 0.001 or > 66% of variation explained.

| | *Risk factor* | *No. Times associated* | *Significance $^\partial$* | *References* |
|---|---|---|---|---|
| *Presence of potential reservoir or intermediate host species* | Species distribution | NA | NA | See *Potential reservoir and intermediate host species* section in Methods |
| *Environmental factors* | Forest cover (tropical forest, evergreen forest, MGVF [+]) | 4 | ***/ **/ **/NS | [16,21,32,39] |
| | Cropland* | 0 | NA | [40] |
| | Cropland to forest ratio[†] | 1 | ** | [39] |
| | Loss of forest cover (forest fragmentation, changes in forest fragmentation) | 3 | **/**/**/**/NS | [21-22,39] |
| | Landscape productivity (i.e. NDVI anomaly, EVI) | 5 | */**/NA/NA/NS/NS/NS | [18,20,41-43] |
| | Distance to rivers[δ] | 0 | NA | [44] |
| | Distance to roads | 2 | * | [20,39] |
| | Human population density[§] | 4 | ***/*/*/** | [20-21, 32,39] |
| *Climatic factors* | Annual temperature range | 2 | **/NA | [16,39] |
| | Annual mean temperature | 3 | ***/*/NS | [16,23,32] |
| | Mean monthly rainfall (rainfall seasonality, evapotranspiration) | 6 | **/**/*/*/NA/NS | [16,18,20,23,32,43] |
| *Bushmeat trade and consumption* | Bushmeat hunting areas | 1 | 0.012 | [16] |
| | Bushmeat trade | | | |
| | Presence of domestic animals[γ] | 0 | NA | [45] |
| | Human population density | | | (see Environmental factors above) |

[+] MGVF: Maximum green vegetation fraction

\* We included crops as several species of bats are known to feed and roost in crop fields.

† [39]. considered the forest cover to cropland ratio. Here we calculated it inversely to reduce the number of No Data cells in the resulting raster, as in our study areas the percentage of cropland is less than 50% of the total area.

NDVI: Normalized Difference Vegetation Index; EVI: Enhanced Vegetation Index.

δ We included distance to rivers as they are an important factor for the presence of bats [46].

§ We included human population density in the environmental factors as it is a prerequisite for spillover from wildlife to humans, even in the absence of activities related to bushmeat consumption.

ᵧ Although domestic animals were not considered as a risk factor in any of the studies of the literature review, it is not clear if some could be implicated in the spillover of EBOV [45].

∂ In some cases the P-value of a risk factor was not reported, or the risk factor was tested but not significant; thus the number of times a risk factor was associated with *Ebolavirus* spillover does not necessarily correspond to the number of values showing significance for that risk factor.

**Generation of weights for risk factors**

We estimated the weight of each risk factor separately for each category: environmental factors, climatic factors, factors related to the bushmeat trade and consumption, and potential reservoir and intermediate host species. We favored this approach because knowledge of how different risk factors from different categories interact with each other, and of the role of these interactions in the spillover of EBOV, is limited. Within each category, except for the potential reservoir and intermediate host species category (see below), pairwise comparisons between factors were done in a five-point scale (i.e., from strongly less important to strongly more important) [47,48]. For the comparisons, we selected studies that aimed to investigate the association of ecological and climatic variables to spillover or outbreaks of Ebola virus. Following the procedure used by Stevens and collaborators [30], we took into account the number of times each factor was tested and its significance (p-value or % of the model explained) in each study, which indicates how strongly a factor is associated to *Ebolavirus* spillover (Table 1). Factors that appeared more frequently and/or with a higher level of significance were considered more important than factors that appeared less frequently or with lower significance [30]. As studies have assessed similar, but not necessarily the same risk factors we considered factors that represent a similar ecological variable (e.g., tropical forest *vs* evergreen forest) as one. Pairwise matrices comparing each factor with each other were built for each category, and factor weights were calculated through an analytical hierarchy process (AHP). This consists in ranking and comparing the importance of each factor regarding all other factors in relation to the outcome [47,48]. We compared pairs of factors within each category (e.g., environmental factors were only compared to the other environmental factors) using a five point-scale to specify the degree of importance of factors. Then a vector of weights is generated that corresponds to the principal eigenvector of the matrix [47].

**Potential reservoir and intermediate host species**

We focused on four groups of species: fruit bats, insectivorous bats, primates and duikers. These were chosen because RNA of *Ebolavirus* or antibodies against *Ebolavirus* have been found in fruit bats and insectivorous bats [49], and bats, primates, and duikers have been implicated in previous *Ebolavirus* spillover events to humans [16]. High mortality rates of primates, especially apes, have also been associated with *Ebolavirus* infection [17,50]. For bats, we took into account all species in which EBOV RNA or antibodies against the virus have been found. For primates, we took into account known species that have been associated with previous spillover events. For duikers, we took into account all species whose distribution maps (www.iucnredlist.org) overlapped with our study areas. In total, we considered the distribution of ten species of fruit bats, four species of insectivorous bats, three species of primates and 12 species of duikers. To assess the relative importance of each species, we assigned a value from 1 to 5 according to the likelihood that the species was a potential reservoir or intermediate host for the virus. We considered that the four groups of species differ in their relevance for spillover, thus we carried out pairwise comparisons between them using a pairwise matrix. Details of species assessments and the species considered in this study are given in S2 Text.

**Creation of suitability maps for spillover**

Four different suitability maps for EBOV spillover were produced, one for each category of risk factors (i.e., environmental, climatic, factors associated with bushmeat trade and hunting, and the potential reservoir or intermediate host species). Each map was produced by combining the different factors within each category. First, each raster layer representing one risk factor was multiplied by its corresponding weight; then the multiplied raster layers within the same category of risk were added to produce a suitability map of their category (Fig 2). On these maps, each pixel represents the suitability of an area of 1 km² for the occurrence of EBOV spillover according to the category of factors. These four maps were standardized to a continuous scale between 0 and 1. To create the final map of suitability for EBOV spillover, the four previously created suitability maps were combined into one by calculating the mean of the four raster layers (Fig 2), considering the four types of factors equally important for the occurrence of spillover.

Following the procedure described above, we created a suitability map for EBOV spillover in Guinée forestière for every month in 2013 to assess seasonal variability in the suitability for EBOV spillover to humans. For these maps, the risk factors that varied on a monthly basis were the NDVI and the mean rainfall; all other factors were considered fixed in time (Table B in S1 Text).

We used the softwares ESRI ArcMap 10.4.1 [51], QGIS 3.10.2 [52] and R 3.5.3 [53] for all data treatment and the creation of maps.

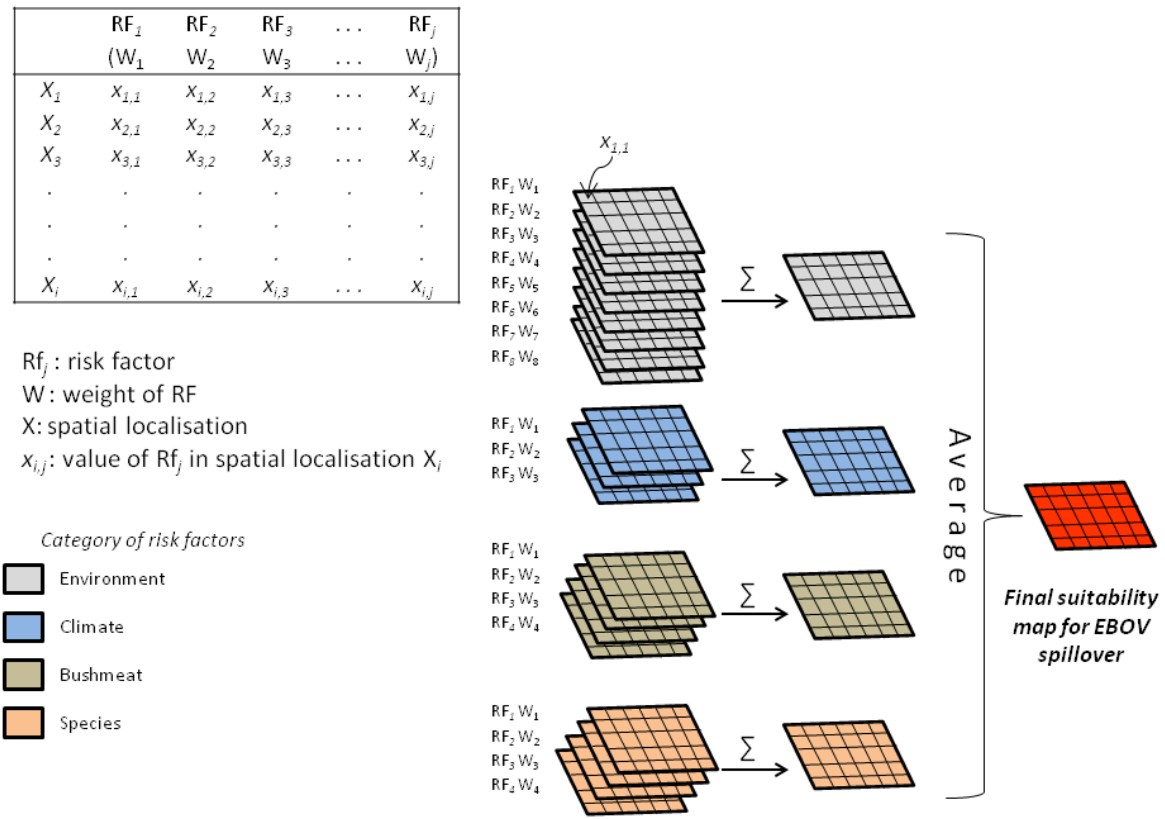

**Fig 2**. Schematic diagram of the GIS-MCE procedure followed to create the suitability maps for EBOV spillover.

**Map validation**

To validate the suitability maps for EBOV spillover, we used eight previously reported spillover events that occurred since 2000 and for which geographic localization was known, and information on the month it took place was available [16,20]. One spillover event was reported in Guinée forestière (December 2013), five in Congo (two in December 2001, one in January 2003 and two in June 2003), and two in Gabon (August 2001). For each of the maps corresponding to the dates of these spillover events, we extracted the pixel value of sites where spillover events were reported. We compared these values with the mean of pixel values of the corresponding full mapped areas. We also calculated the mean pixel value of a buffer area of 5 km and 10 km around spillover sites to account for movements of potential reservoir or intermediate host species and humans.

Furthermore, as another way to assess the performance of the method, we created suitability maps of *Ebolavirus* maintenance in fruit bats and insectivorous bats in Guinée forestière. Details on the creation of these maps are provided in S3 Text. It remains unclear which factors are relevant for the maintenance of the virus in bat populations [11,49], thus we considered the same risk factors as for the spillover maps, but left out of the model the factors related to the transmission to humans (i.e., factors associated with the trade and consumption of bushmeat, and the human population density). Weights for each risk factor were estimated within their corresponding category (i.e., environmental, climatic and bat species), and pairwise comparison matrices were built for each category of risk factors. The suitability maps for *Ebolavirus* maintenance corresponding to the three categories of risk factors then were produced. Finally, these three maps were combined, by calculating the mean, to create maps of *Ebolavirus* maintenance for December 2016 and for March 2017. These maps were compared to data of seroprevalence for *Ebolavirus* in bats that were sampled in the area in December 2016 and March 2017 [12]. We examined if the average suitability index of sites where at least one bat was positive to *Ebolavirus* antibodies was higher than that of sites where all bats were negative using a one-sided two-sample test [54].

**Uncertainty and sensitivity analysis**

To assess the sensitivity of the method to the weights assigned to each risk factor, we carried out a sensitivity analysis separately for each category of risk factors. For this, we increased or decreased the original weight of each factor by 25% while proportionally adjusting the weights of the other factors ($w_i$ in Equation 1) within its category so that the final sum of weights equaled 1.

$$w_i = (1 - w_m) * \frac{w_{i0}}{1 - w_{m0}}, 1 \leq i < n, i \neq m \tag{1}$$

where $w_m$ is the changing factor, and $w_{m0}$ and $w_{i0}$ are the weights of the changing factor and the $i$-th factor in the base model, respectively.

Each of the newly calculated weights combination was incorporated into the GIS-MCE to produce a suitability map for spillover in our three study areas according to the four categories of risk factors. This resulted in 16 environmental, six climatic, eight bushmeat-related and eight species-related suitability maps for each study area.

To assess the contribution of the variation weights of each factor on the variation of the suitability map of their corresponding category, we calculated *delta* for each +/- 25% variation introduced to the weight of each risk factor using the following equation:

$$delta = \frac{abs(x_f - x_i)}{x_f} \tag{2}$$

where $x_f$ is the pixel value in the suitability map of category $f$ of risk factors and $x_i$ is the pixel value in the output map when a variation in weight is introduced to risk factor $i$. Then we calculated the mean pixel value and the standard deviation of the output maps for each 25% variation of weight.

Additionally, we produced an uncertainty surface for our three study areas. An uncertainty surface shows the variability in the outcome maps when changes are introduced in the risk factors. It evaluates how uncertainties in the input translate to uncertainty in the outcome. To create the uncertainty surface, we followed the same procedure as above, but introduced a ± 25% weight variation at the category level instead of at the individual risk factor level. This resulted in eight (two per category) new maps of suitability for EBOV spillover for each area. We estimated *delta* using equation 2; where $x_f$ represents the pixel value in the EBOV spillover suitability map, and $x_i$ the pixel value in the output map of category *i* of risk factors. A map of the mean, which represents the average relative change per pixel, and a map of the standard deviation, were produced.

## Results

**Weights of risk factors**

Table 2 shows the weights estimated from the review of the literature for the risk factors of each category (environmental, climatic, factors associated with bushmeat trade and the four groups of host species). The pairwise comparison matrices for the AHP are shown in the S4 Text (Tables A-D in S4 Text). Human population density may be a risk factor even in the absence of activities associated with bushmeat trade and consumption, and therefore was considered as a risk factor in the environmental category. Forest cover and forest cover loss had the highest weights among the environmental risk factors. Of the climatic risk factors, monthly rainfall contributed over 68% of the total weight in the category. Bushmeat hunting areas and areas of trade were considered equally important for EBOV virus transmission and more important than the human population density and the presence of domestic animals. The weights of fruit bats and insectivorous bats were equal to and higher than those of duikers and primates.

**Table 2.** Weights for risk factors resulting from the AHP.

| Risk Factor | Weight |
|---|---|
| *Environment* | |
| Forest cover | 0.255 |
| Cropland | 0.048 |
| Cropland : forest ratio | 0.096 |
| Loss of forest cover | 0.255 |
| Landscape productivity | 0.096 |
| Proximity to rivers | 0.032 |
| Proximity to roads | 0.05 |
| Human population density | 0.167 |
| *Climate* | |
| Annual temperature range | 0.211 |
| Mean annual temperature | 0.102 |
| Mean monthly rainfall | 0.686 |
| *Bushmeat* | |
| Hunting areas | 0.380 |
| Bushmeat trade | 0.380 |
| Human population density | 0.179 |
| Presence of domestic animals | 0.062 |
| *Species* | |
| Fruit bats | 0.375 |
| Insectivorous bats | 0.375 |
| Duikers | 0.125 |
| Primates | 0.125 |

## Suitability maps for ebolavirus spillover

Maps for each category of risk factor in the three study areas are shown in Fig 3. Suitability for EBOV spillover due to environmental risk factors was more variable in space in Guinée forestière than in the study areas in Congo and Gabon. In Guinée forestière, most of the area shows intermediate to high suitability for EBOV spillover for this category, whereas in Congo and Gabon only some small patches show high suitability for EBOV spillover. The maps of climatic risk factors show that in the three study areas, there is a spatial gradient of suitability with clearly localized areas of high suitability. The maps of suitability for EBOV spillover associated with bushmeat risk factors are quite different among the three study areas. In Guinea, most of the study area has intermediate values with patches of higher suitability. The study area in Congo also shows patchiness, but the northern part is in general more suitable for spillover compared to the rest of the mapped area. In Gabon, three areas of higher suitability can be distinguished in the northeast, southeast and western regions. Maps of suitability given by the potential reservoir and intermediate host species show that areas with higher suitability for EBOV spillover are found in the central region in Guinea, towards the northwest in Congo and in a small area in the southwest in Gabon.

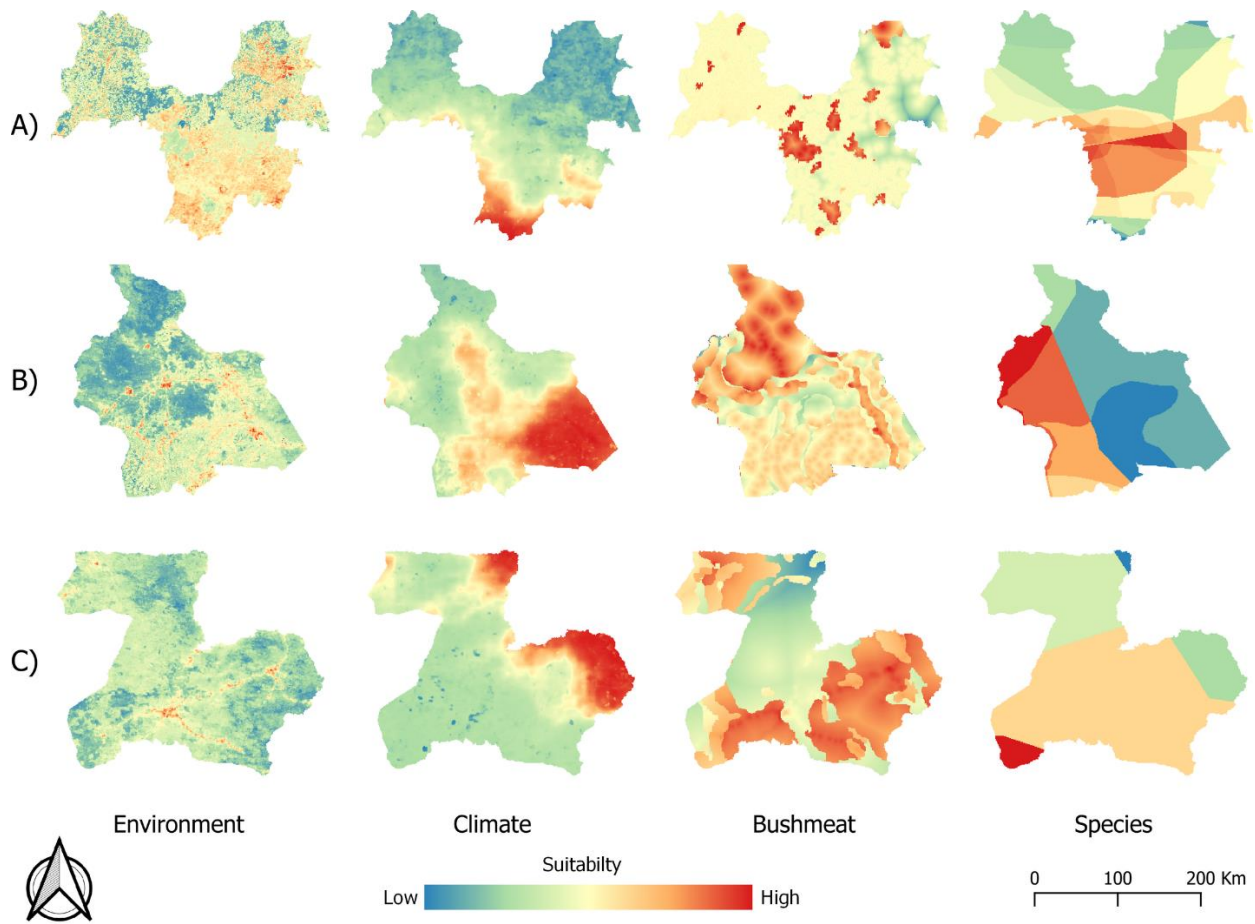

**Fig 3**. Maps for the four categories of risk factors in study areas in A) Guinea (December 2013), B) Congo (December 2001) and C) Gabon (August 2001) (dataset available at https://doi:10.18167/DVN1/FZANMS).

Environmental, climatic, bushmeat and species maps were combined to create the final suitability maps for EBOV spillover in our three study areas (Fig 4). These maps show that there is high spatial variability in the three study areas. In Guinée forestière, areas that are more suitable are located in the southern half, with small patches scattered within this area showing the highest suitability for spillover. In the area in Congo, suitable areas differ in relation to date, but in general, the western part is more suitable for spillover than the eastern part, especially during the short dry season (December-January). Nevertheless, in December 2001 the southeast comes out as a suitable area for spillover as well. Finally, in Gabon there are less areas of very high suitability for spillover, with suitability being in general higher towards the east and south of the mapped area.

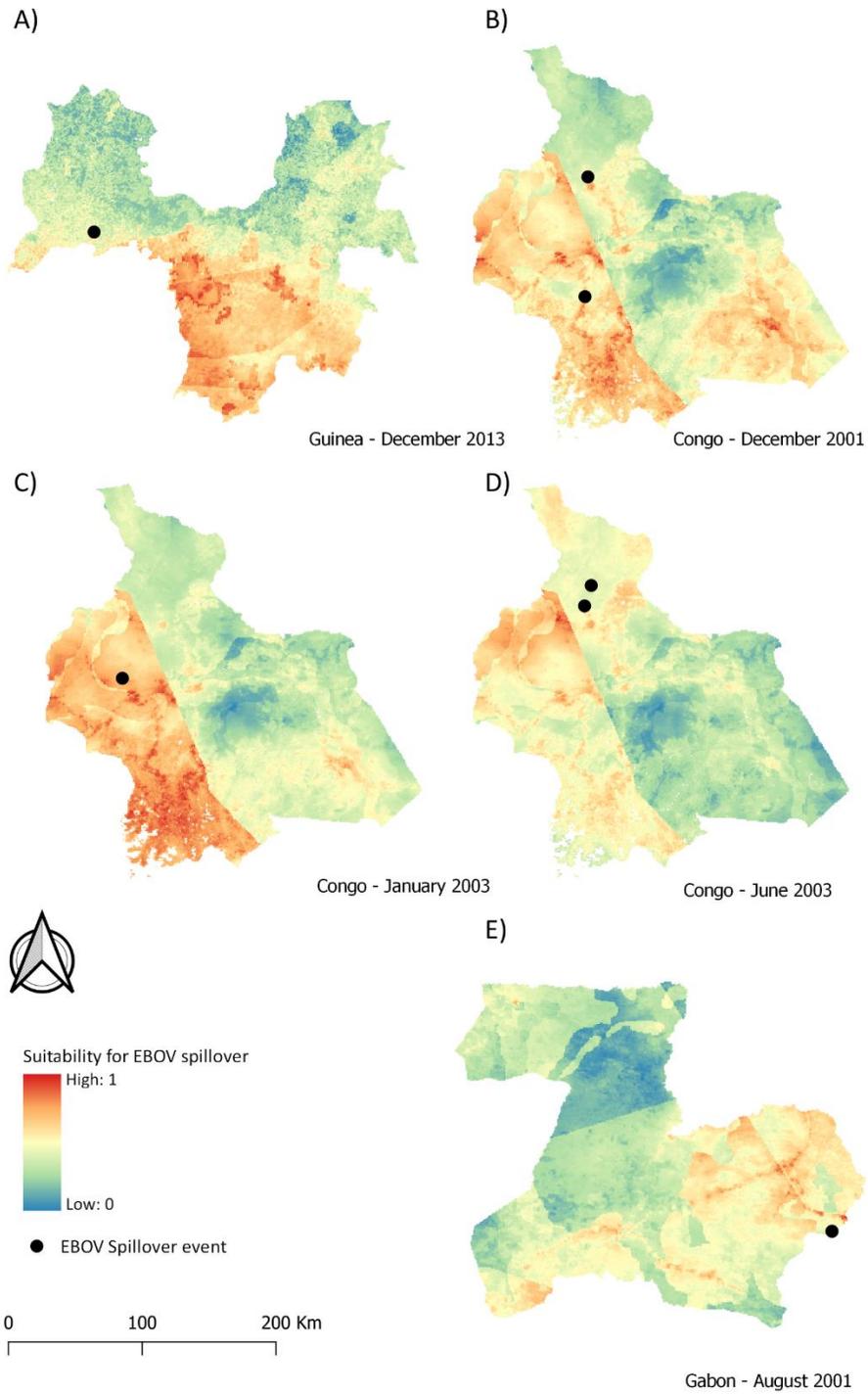

**Fig 4**. Suitability maps for EBOV spillover in areas in Guinea (A), Congo (B-D) and Gabon (E) (dataset available at https://doi:10.18167/DVN1/FZANMS).

The resulting maps highlight a strong temporal variation in suitability for EBOV spillover. For instance, the maps in Congo at three different dates (Fig 4) show that although the western area has consistently higher suitability, other areas such as the southeast were more suitable for spillover in December 2001 and unsuitable in June 2003. In contrast, the northern part was unsuitable for spillover in December 2001, and transitioned to higher suitability in June 2003. Temporal variation is also observed within a year in Guinée forestière (Fig 5). Our maps show that although the central region remained an area suitable for EBOV spillover throughout the year, the suitable area decreased as the rainy season progressed, and then started to increase again from August on. Other areas also show a marked variability. For instance, the northwest region was unsuitable for spillover for most of the year, but became suitable in May, June and October; and areas in the northeast, unsuitable for most of the year, showed a high suitability in June. The pixel value where the village of Meliandou, site of the index case of the Ebola outbreak in West Africa, is located ranged from 0.412 to 0.689, with an average annual suitability of $0.522 \pm 0.083$ (SD). In a random sample of 500 points, average annual suitability ranged from 0.079 to 0.874, with standard deviations ranging from 0.038 to 0.135.

**Map validation**

Epidemiological data to validate our spillover maps are scarce; we therefore used the localization of spillover events previously reported in the mapped areas. Pixel values of the location of each spillover event (Table 3 and Fig 6) show that in Congo, all but one of the spillover events fell in areas on the upper half of the distribution, with three of the five spillover events being located on the upper quartile. The spillover events reported in Guinea fell just over the second quartile, whereas those in Gabon took place in areas on the lower half of the distribution. The average pixel value of sites where a spillover event was located was $0.502 \pm 0.135$ (SD). The mean pixel value of a 5 km or 10 km buffer around spillover sites tended to be higher than the raster mean pixel value for Gabon and for three spillover events in Congo (December 2001, January 2003 and June 2003), whereas it was similar for Guinea and for two spillover events in Congo (December 2001 and June 2003) (Table 3).

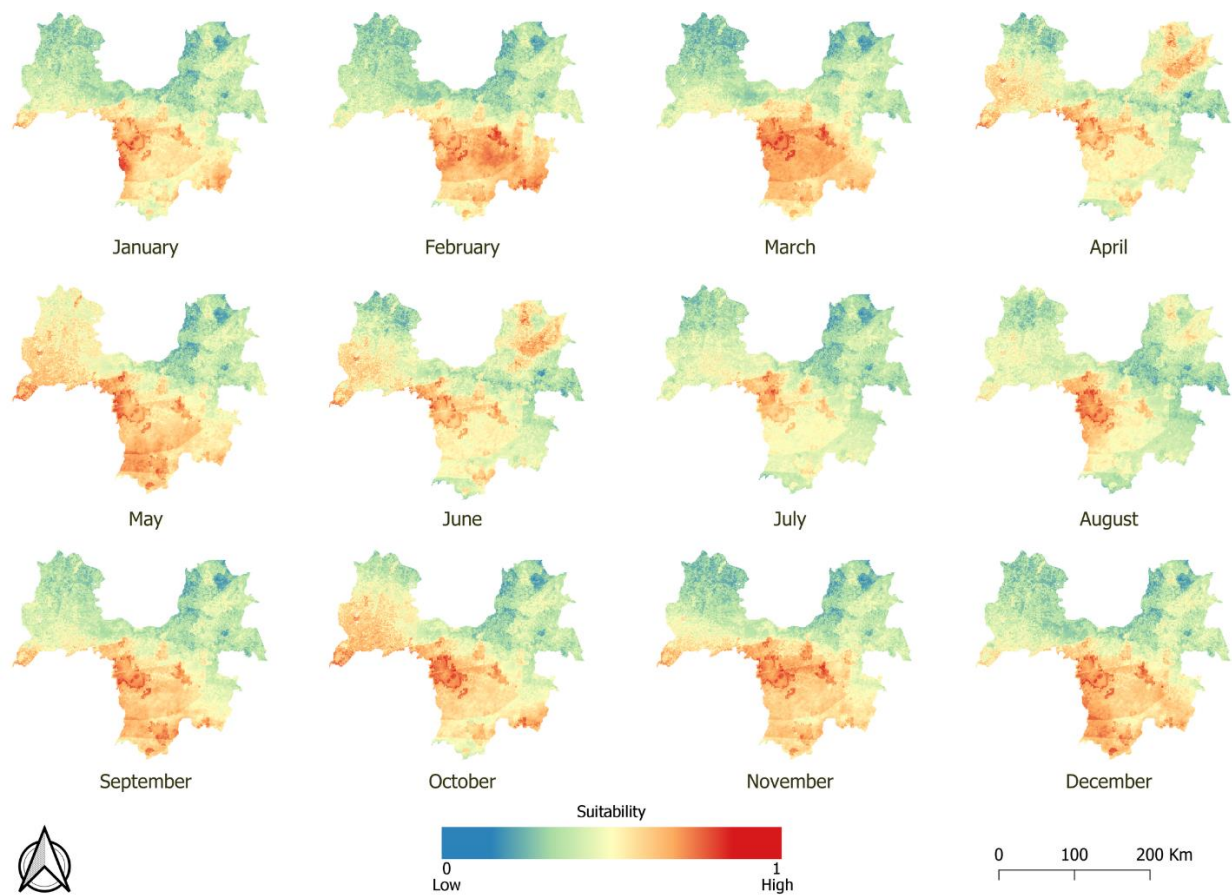

**Fig 5**. Suitability maps for EBOV spillover in forested Guinea in 2013 (dataset available at https://doi:10.18167/DVN1/FZANMS).

**Table 3.** Ebolavirus spillover suitability index of sites where spillover events have been previously reported compared to the values (mean and standard deviation) of the whole study area. The percentile of suitability index for the spillover site is shown.

| Country | Date | Spillover site | | Buffered area around spillover site | | Study area |
|---|---|---|---|---|---|---|
| | | *Spillover* | *Percentile* | *5km* | *10 km* | *Mean ± SD* |
| *Guinea* | December 2013 | 0.451 | 0.52 | 0.438 | 0.438 | 0.459 ± 0.192 |
| *Congo* | December 2001 | 0.668 | 0.64 | 0.531 | 0.537 | 0.485 ± 0.165 |
| | December 2001 | 0.566 | 0.87 | 0.486 | 0.474 | |
| | January 2003 | 0.675 | 0.77 | 0.690 | 0.690 | 0.492 ± 0.189 |
| | June 2003 | 0.415 | 0.48 | 0.421 | 0.429 | 0.415 ± 0.148 |
| | June 2003 | 0.553 | 0.83 | 0.481 | 0.470 | |
| *Gabon* | August 2001 | 0.338 | 0.39 | 0.405 | 0.459 | 0.387 ± 0.152 |
| | August 2001 | 0.338 | 0.39 | 0.405 | 0.459 | |

**Ebolavirus maintenance in bats.** The northwestern and central regions of Guinée forestière showed intermediate to high suitability for *Ebolavirus* maintenance in fruit bats in December 2016 and in insectivorous bats in March 2017 (Fig 7). This result is consistent in suitability maps that considered fruit bats and insectivorous bats together (Fig A in S5 Text). Pixel values of sites where at least one bat tested positive for *Ebolavirus* antibodies are higher than pixel values of sites where all sampled bats came out negative, particularly for insectivorous bats (Table 4 and Fig 8). This difference tended to be significant when comparing the means of all sampled sites irrespective of the type of bat (p = 0.09).

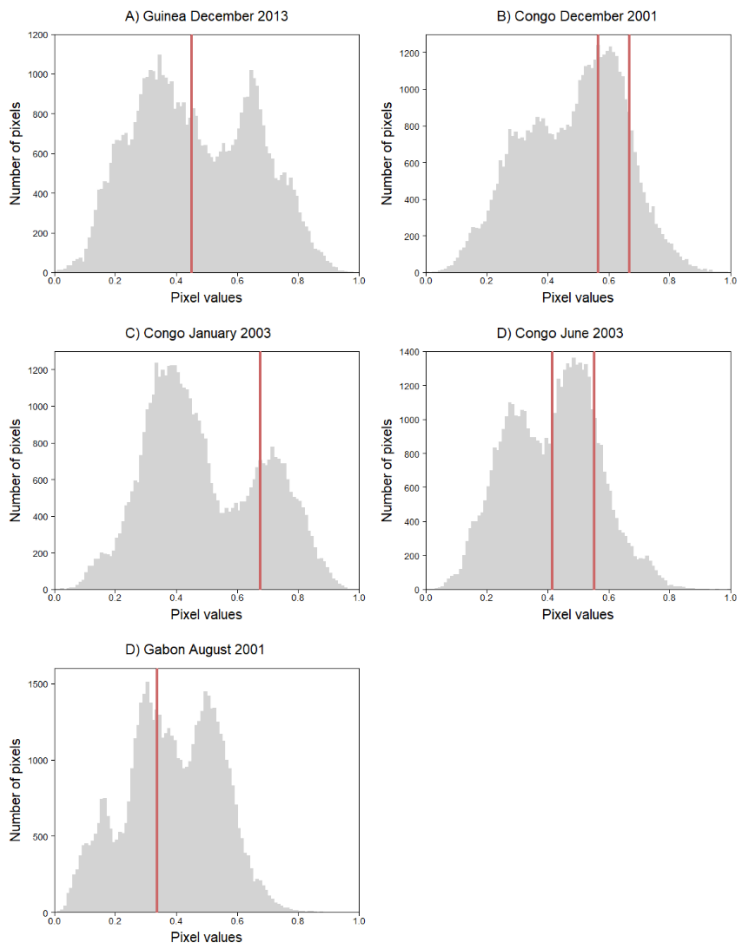

**Fig 6.** Pixel values distribution in suitability maps for EBOV spillover in Guinea, Congo and Gabon. Red lines show pixel values of sites where EBOV spillover were located.

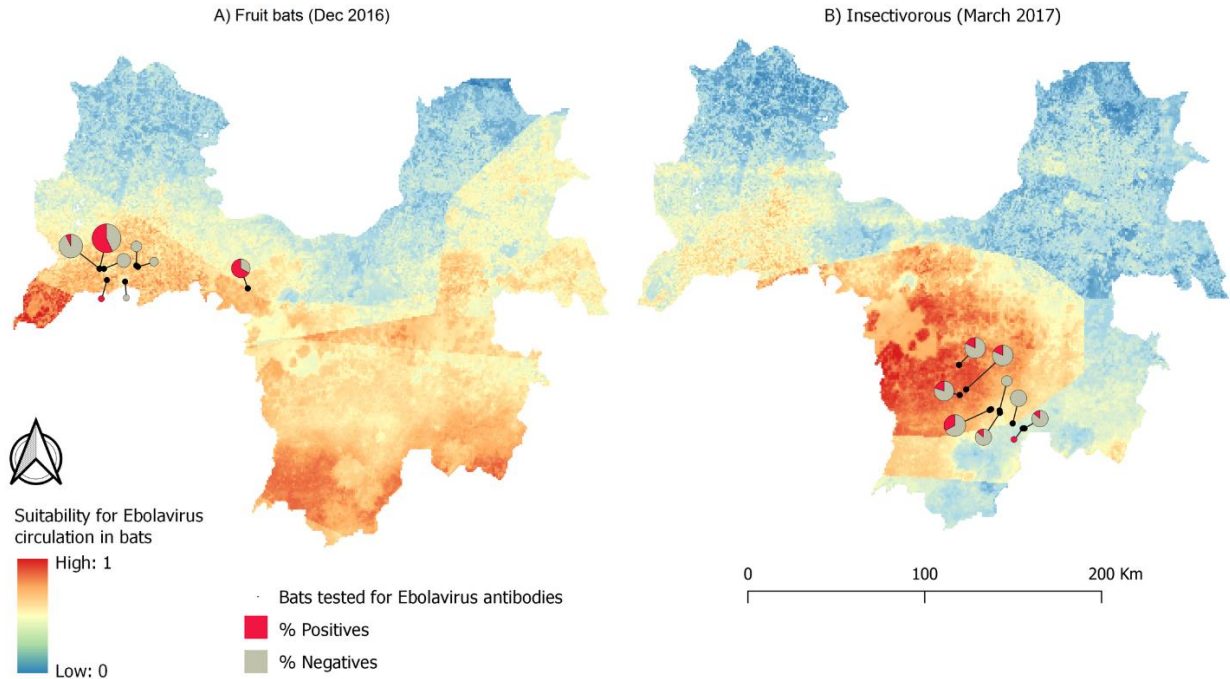

**Fig 7.** Suitability maps of *Ebolavirus* maintenance in fruit bats (A) and insectivorous bats (B) in Guinée forestière. The size of the pie charts is scaled to sampling effort in each site (i.e. n=1 for the smallest pie chart; n=21 for the largest pie chart) (dataset available at https://doi:10.18167/DVN1/FZANMS).

**Table 4**. Mean ± standard deviation of suitability index in sampling sites where at least one bat tested positive for *Ebolavirus* antibodies and where all bats tested negative. * Significance at $p < 0.1$ for t-test.

|  | **Bats tested for EBOV antibodies** | |
| --- | --- | --- |
|  | *At least one positive* | *All negative* |
| All bats* | 0.670 ± 0.123 | 0.613 ± 0.057 |
| Fruit bats | 0.625 ± 0.044 | 0.612 ± 0.060 |
| Insectivorous bats | 0.620 ± 0.189 | 0.570 ± 0.008 |

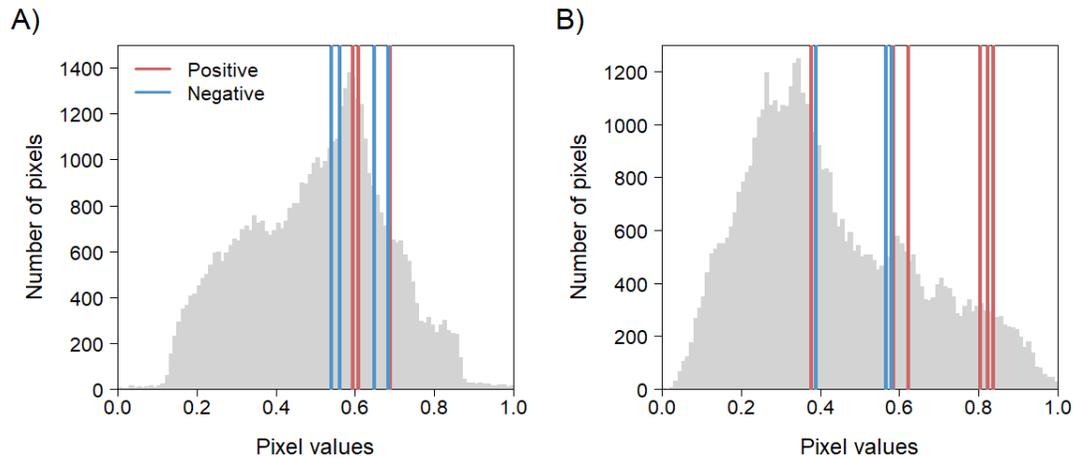

**Fig 8**. Pixel values distribution in suitability maps for *Ebolavirus* circulation in fruit bats (A) and insectivorous bats (B). Lines show pixel values of sampling sites where at least one bat tested positive for *Ebolavirus* antibodies (red lines) or where all bats tested negative (blue lines).

**Uncertainty and sensitivity analysis.** The uncertainty surface shows that the introduced variation of 25% to the weight of the four categories of risk factors has a small effect on the pixel value of the EBOV suitability maps, with a maximum standard deviation of the mean relative change of less than 0.07 (Fig 9). This indicates that the suitable areas for EBOV spillover are robust and remain stable when the weights of the four types of risk factors are varied ± 25% of their original weight.

Fig 10 shows the results of the sensitivity analysis within each category of risk factors for the three study areas for a weight variation of ± 25%. In the environmental category, forest and forest loss were the risk factors that contributed the most to the variation in our three study areas, followed by landscape productivity (i.e., NDVI) and human population density. However, there is variation on the amount each of these factors contributed to the variation in suitability among countries. Loss of forest cover was the most sensitive parameter in the environmental category in Congo, whereas in Guinea and Gabon, it showed similar sensitivity to forest cover. Changes to the weight assigned to human population density contributed more to the variation in pixel value in Guinea than in Congo and Gabon, whereas landscape productivity was the third most sensitive risk factor in Congo and Gabon, and the fourth in Guinea.

Variation introduced to the weight of crops, crops to forest ratio, proximity to roads and proximity to rivers contributed little to the variation of the environmental map of suitability in the three areas studied.

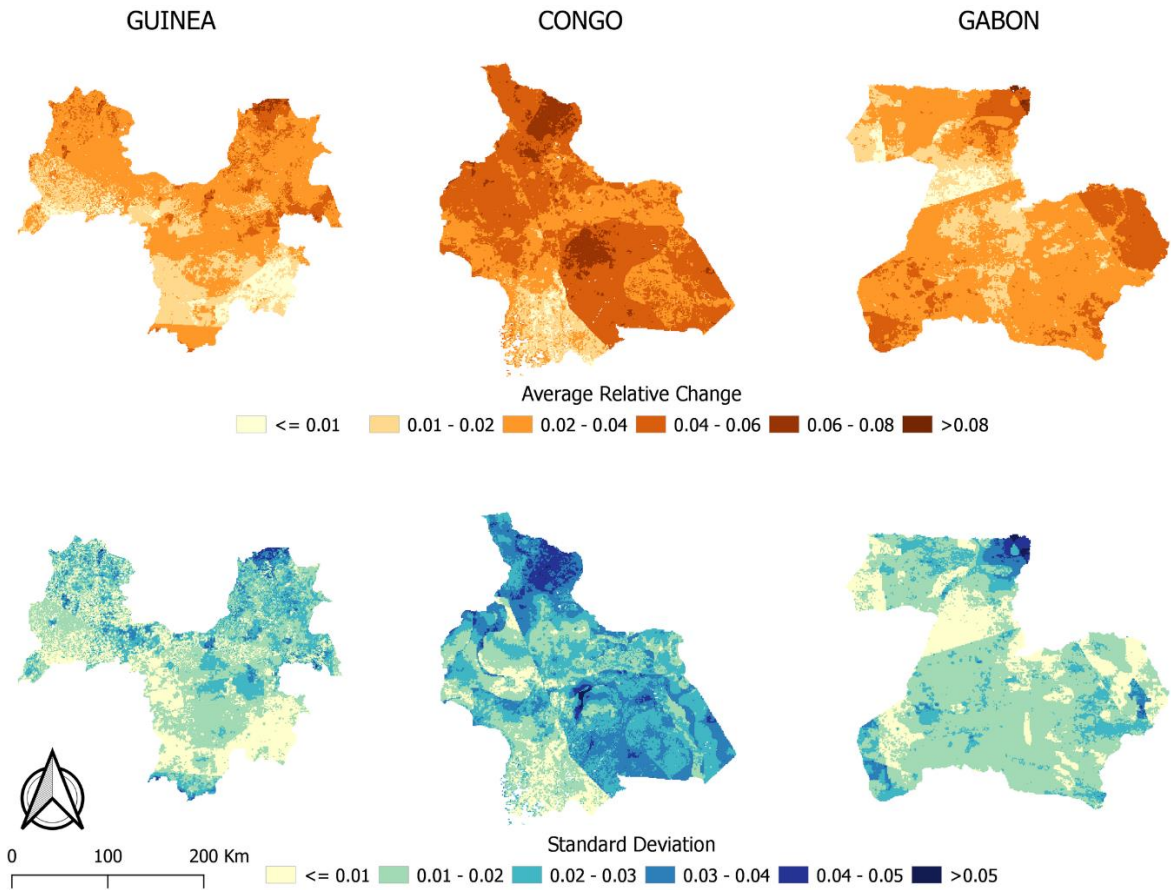

**Fig 9.** Average relative change and standard deviation of suitability maps for EBOV spillover in the study areas in Guinea, Congo (December 2001) and Gabon (dataset available at https://doi:10.18167/DVN1/FZANMS).

Among the climatic risk factors, monthly rainfall was the most sensitive parameter in the study areas of the three countries, and was twice as sensitive in Guinea and Gabon compared to Congo. Variation on the weight of annual temperature range and of annual mean temperature had a smaller impact on the climatic variability in the three areas, with the latter being the least sensitive of the climatic risk factors.

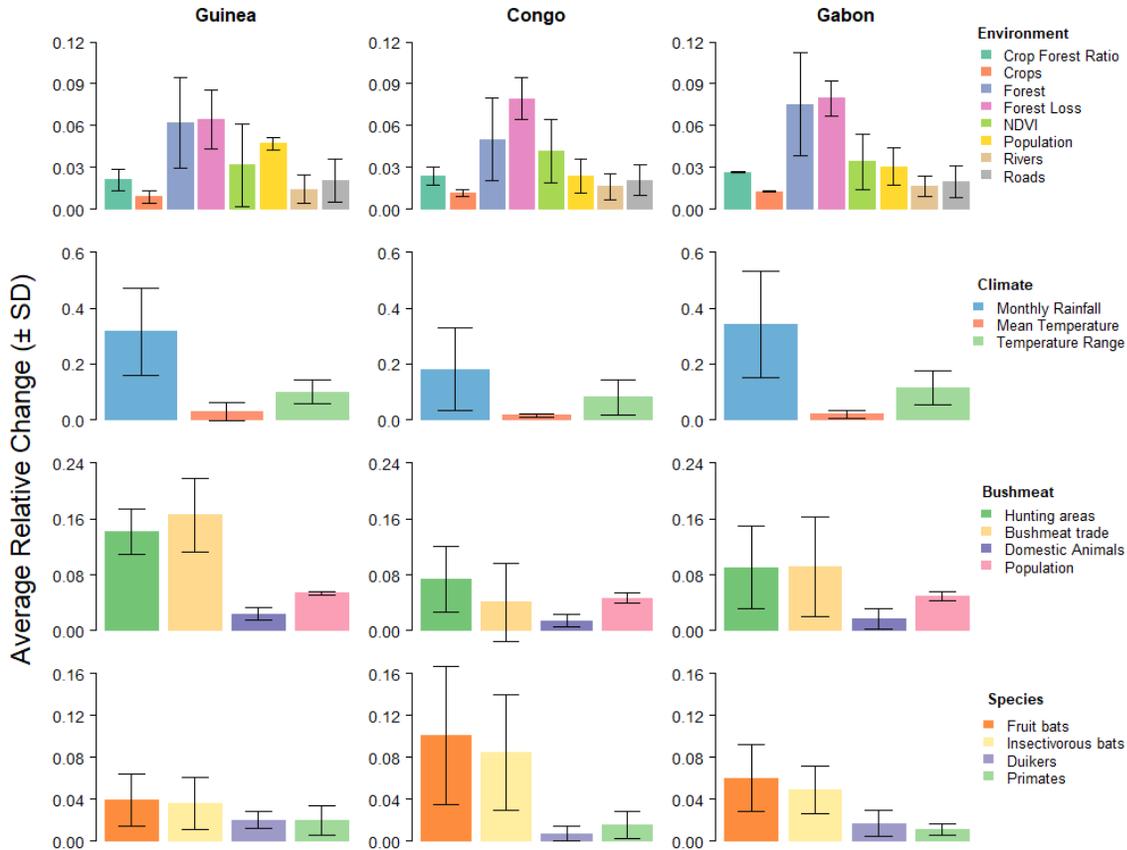

**Fig 10**. Average relative change in pixel value due to a ± 25% change introduced to the weight of risk factors in each category of risk factors (*i.e.* environmental, climatic, associated with bushmeat consumption and associated with potential reservoir/host species) in the three studied areas. Colors correspond to risk factors in the four categories.

Bushmeat hunting areas and bushmeat trade contributed the most to the variation in the bushmeat suitability map in the three study areas, followed by human population density. The average relative change in pixel value due to the variation in weight of hunting areas and hunting trade was more than twice that of domestic animals and human population density in Guinea, but this difference was less marked in Congo and Gabon.

Finally, fruit bats and insectivorous bats were clearly more sensitive than primates and duikers in Congo. This difference was less obvious in Gabon, whereas in Guinea the four groups of species had a similar contribution to the overall variation in the suitability map of species.

## Discussion

In this study, we used the GIS-MCE approach to produce maps at the landscape scale in Guinea, Congo and Gabon. Our maps show that suitability for EBOV spillover to humans was variable in space in the three mapped regions, with previous spillover events being located in areas of intermediate suitability, with higher suitability values for our study area in Congo. Our results also show strong variation in the suitability for EBOV at different times of the year. The sensitivity analysis shows that the maps produced are robust, supporting our choice of risk factors for EBOV transmission from animals to humans.

The 2014 Ebola outbreak in West Africa made clear that areas previously considered free from the risk of an epidemic might be exposed to future outbreaks. Numerous studies have tried to understand what led to this outbreak. Genetic studies suggest that the virus was introduced from Central Africa within the previous decade [6]. Antibodies against EBOV found in one person in Guinea in 2012 [5] suggest that the virus might have already been circulating in the human population two years before the outbreak. Given that antibodies against the virus can last at least 40 years in one person [55], it is also possible that the positive individual acquired the infection elsewhere several years ago. Similarly, the virus could have come from Central to West Africa in a reservoir host. Gire et al. [6] point to a single spillover event from a reservoir host, plausibly insectivorous bats [56], to the human population.

Previous studies have used ecological niche modeling to map the risk of Ebola virus spillover at a national or a continental scale [16,18,19,23]. The GIS-MCE approach used here allows integrating a large amount of different types of information in a simple procedure. It may overcome some of the limitations of niche ecological models such as not accounting for or misrepresenting the interaction between multiple hosts, and not accounting for other elements of the system such as behavior of hosts or demography [57]. Also, risk factors, their relative importance and their association to spillover may differ at different spatial scales [58]. Here we produced maps at a finer scale, which may help to identify which local factors are more important for the spillover of the virus at distinct ecological regions. These regional landscape models allow to better account for local differences in risk factors and may be more adequate for surveillance in specific regions.

The GIS-MCE approach has proven useful in risk mapping diseases when not enough outbreak data exists. This method has been applied to map the risk of transmission of several vector-borne diseases such as malaria [29,59], Rift Valley fever [26], and visceral leishmaniasis [60]. However, to our knowledge, this is the first time that it has been applied to Ebola virus disease, a disease with a relatively unknown transmission cycle.

**Classification of risk factors**

The main factors for the occurrence of EBOV spillover to humans are still unknown, and some factors may be more important than others. For example, some models consider the distribution of reservoir or host species as more important than environmental factors for determining favorable areas for Ebola virus occurrence [39], whereas others have considered environmental, climatic and anthropogenic factors as more relevant for spillover [18,20,32]. In addition, the relative importance of the different types of factors may vary according to a given context or species. For example, bushmeat hunting of primates and duikers could be a more important risk factor in areas with a dense forest cover, such as Congo and Gabon, compared to more fragmented natural areas, such as Guinée forestière, where these animal populations could have been already depleted or extinguished. In contrast, in landscapes with a mixture of crops and forest, some bat species, more adapted to fragmented landscapes, could pose a risk of spillover when hunted as bushmeat [15], and through indirect contact with their feces or saliva present in the environment [40]. Indeed, numerous EBOV spillover events in the Congo basin have been associated with carcasses of primates, duikers and bats [16], whereas in Guinée forestière, the patient zero of the 2014 outbreak was identified as a two-year old boy who plausibly came into contact with live insectivorous bats [56]. There are clear differences in terms of landscape between our study areas, mainly between Guinée forestière compared to the study areas in Congo and Gabon. In the former, pixels where forest represented 50% or more of the pixel area covered only 44%, and 14% of the study area was covered by crops, whereas in Congo and Gabon forest represented over 60% of the total area, and crops represented only 2.1 and 0.02% of the area, respectively. Thus, it is possible that the risk factors and/or their relative importance for a spillover differ between regions. In this study, we decided to give the same weight to the four types of risk factors (environmental, climatic, factors associated with bushmeat consumption, and the potential reservoir and host species) in our three study areas because the relatively scarcity of known EBOV spillover events and the gaps in our understanding concerning spillover factors prevented a more in-depth study of potential differences in risk factors between Central and West Africa. Nevertheless, as more data and knowledge becomes available, models can be refined and used for specific regions.

**Suitability for EBOV spillover**

Suitable areas for EBOV in Guinée forestière were located in the central-western area. The presence of the hammer-headed bat (*Hypsignathus monstrosus*), the Angolan fruit bat (*Lissonycteris angolensis*), and the greater long-fingered bat (*Miniopterus inflatus*) in this area, but not in the most northern regions, partly explains this result. Other factors that could have influenced a higher suitability in the southern half of Guinée forestière are the cropland to forest ratio, which is higher in the southern half, and bushmeat hunting areas, which are larger and preferentially located in the central-southern region (Fig B in S1 Text).

In Congo, the western part of the study area showed high suitability for EBOV spillover at the three dates mapped. This results in part from the presence of the Egyptian fruit bat (*Rousettus aegyptiacus*) which, according to the IUCN distribution maps used, is not present in the eastern part. Experimental inoculation studies involving this bat species have not found clear evidence of EBOV viral replication and shedding, and only low levels of viral RNA in tissues, suggesting it is an unlikely reservoir for EBOV [61,62]. We included this species in our study because antibodies against the virus have been found in individual bats [12,49]. Its presence also is favored by human activities linked to deforestation within favorable areas for EBOV [63]. Land cover type and climatic factors such as temperature seasonality appear to be important in determining habitat suitability for potential reservoir bat species in Africa [64]. More studies are needed to disentangle what factors are important in the distribution of bats and their role, if any, in EBOV maintenance and spillover. Forest cover and climatic factors also contributed to higher suitability in certain areas on the eastern part of the study area in Congo (Fig C in S1 Text). In Gabon, suitable areas were located on the southeastern part of the study area, which also is the region bordering the study area in Congo. This area showed higher suitability in risk factors associated with climate and bushmeat (Fig 3C). Our study areas in Congo and Gabon border each other and they are similar in terms of landscape, population density and climate. The differences observed on the maps between these two regions are likely due to temporal variability; the maps for Congo do not correspond to the same date as the map for Gabon.

**Temporal variability in suitability for EBOV spillover**

Temporal variation in suitability for EBOV spillover was present in the maps of Guinea for 2013 (Fig 5) and in the maps of Congo for 2001 and 2003 (Fig 4). This variation is due to the only two factors included in our model that varied on a monthly basis: landscape productivity (i.e. NDVI) and precipitation. In fact, rainfall maps closely reflect climate maps (see rainfall maps in Figs B-D in S1 Text, and climate maps in Fig 3 and Fig B in S5 Text). The increase in suitability in Guinea, in the northwest in May and October, and in the northeast in June, was greatly due to an increase in precipitation in these areas during those months (Fig C in S5 Text), illustrating the relevance of rainfall in our models.

The transition from the dry to the rainy season has been identified as favorable for *Ebolavirus* spillover [20,41,42]. Seasonal climatic variability can be related to the risk of spillover to humans through factors that increase the likelihood of contact between maintenance, intermediate and target hosts on the one hand, and through factors that affect virus circulation and shedding in the maintenance host on the other. During the rainy season, environmental conditions might become more favorable for encounters between a reservoir and a host. For example, fruit trees, such as mangoes and figs, on which bats, other mammals and humans feed, produce fruit during this season. An increase in competition for such resources may consequently lead to more contacts (direct or indirect) between host species and humans at this time of the year. In fact, *Ebolavirus* spillover has been associated with plant phenology [43], indicating that the timing of flowering and fruiting of certain plant species is a factor that can favor *Ebolavirus* spillover events. However, the diversity of fruit-producing plants in these forest habitats and their asynchrony in terms of their fruiting period make it difficult to provide clear spatial and seasonal patterns of *Ebolavirus* risk.

Human behaviors, notably those related to bushmeat consumption, also change according to seasons. In sub-Saharan Africa, bushmeat consumption can increase when access to other food resources diminishes, such as during the dry season, when fish are unavailable, or during the lean season (i.e., the period between planting and harvesting) when rural families' incomes drop [65]. In Guinée forestière, although hunting is only allowed from December to April, bushmeat is hunted and consumed all year round [66]. In this region, the greater cane rat seems to be the most consumed wild mammal, but it is hunted mainly during the rainy season, whereas bats, although apparently less preferred, are hunted throughout the year [67].

Other factors that can play a role in the temporal variability of the risk of spillover are those associated with differences in immunity in the maintenance host at different times of the year. Seasonal climatic variation can affect virus circulation and shedding in maintenance host species, and thus spillover probability [68,69]. For example, the viral load in bats can increase when the immune system is compromised, such as during reproduction or during periods of food shortage [68,70]), and an association between seasonal spillover events and seasonal reproduction in bats has been found for another filovirus, the Marburg virus [71]. Moreover, seasonality in virus circulation and shedding in the maintenance host might differ among species and between different regions. For example, the straw-colored fruit bat breeds once a year, and its gestation period seems to last longer in Guinea than in Central Africa [72]. In contrast, the hammer-headed bat breeds twice a year, with the males grouping into 'leks' to attract females [72,73]. All this adds to the complexity of modelling the risk of Ebola virus transmission from wildlife to humans.

**Validation**

Data that could potentially validate our maps are limited due to the small number of EBOV spillover events that have occurred, particularly in Guinée forestière, where only one case of spillover has been recorded. The need of spillover events localised in space and time (i.e., month when the spillover took place) further reduced the data available for validation. Moreover, the lack of data for some of the risk factors before the year 2000 prevented us from using spillover events reported before that date.

In Guinea and Gabon, the pixel value of EBOV spillover sites was close to the mean pixel value of the whole study area, whereas in Congo, it was higher for four of the five spillover events (Table 3). When taking the mean pixel value of a buffer area around the spillover sites, three out of five of the spillover events in Congo and the two spillover events in Gabon were found in areas of higher suitability compared to the mean pixel value of their corresponding study area. One pixel in our maps represented an area of approximately 1 km$^2$, but suitable areas for virus transmission are probably larger, since the animal species carrying the virus may travel several kilometers. For example, African straw-colored fruit bats (*Eidolon helvum*) can travel over 35 km from their roosting site during the rainy season, when food availability is high, and over 90 km during the dry season [74]. In West Africa, tagged bats of this species have been shown to travel distances of over 500 km (J. Cappelle pers. comm.). Primates as well as humans can travel several kilometers when foraging [75]. Thus, a buffer area around the spillover site may be more accurate in representing suitable areas for spillover while at the same time accounting for the spatial variability at a resolution of 1 km$^2$.

In general, spillover events in Guinea and Gabon were located in areas of intermediate suitability, whereas in Congo, three of the five spillover events were located in areas of higher suitability. Different mechanisms of virus transmission may operate in areas of high *vs* areas of intermediate suitability. For instance, a previous study found that the geographical location of Ebola virus in mammals, including humans, was positively correlated with human population density and distance to roads in areas of intermediate favorability, whereas forest cover and mosaics of forest and crops were more important in areas of higher suitability [39]. Other studies also found that *Ebolavirus* spillover was positively associated with forest cover loss and forest fragmentation [21,22]. A recently fragmented landscape may affect wildlife behavior and favor more anthropophilic species. This can result in different community composition and species abundances (e.g. [76]), and consequently in a different risk of disease spillover. Differences in landscape, climate and mammal assemblages between Guinée forestière, Congo and Gabon may influence the mechanisms or factors involved in an EBOV spillover event (see discussion on Sensitivity Analysis below).

Data derived from a survey of Ebola viruses in bats in Africa [12] allowed us to assess the pertinence of our maps of *Ebolavirus* maintenance in bats. Areas where bats tested positive for *Ebolavirus* antibodies had higher pixel values than those where all of the bats tested negative. This difference was significant at $p < 0.1$, suggesting that the method correctly identifies suitable areas for virus maintenance. Nevertheless, this result must be interpreted cautiously as the presence of antibodies can reflect a past infection that could have been acquired elsewhere. Moreover, fewer bats were caught in areas where all bats came out negative, and there were differences in the number of insectivorous and fruit bats sampled at the two different dates. Insectivorous bats were underrepresented in sampling sites in December 2016 (2 out of 58), and fruit bats were underrepresented in March 2017 (2 out of 71), confounding the effects of time and space of *Ebolavirus* maintenance in both groups. It is possible, at least for fruit bats, that the difference in the number of bats sampled between December and March was related to seasonal migration [77,78]. A longitudinal study on bats currently carried out by members of our group in Guinée forestière will inform us about changes in the bat community throughout the year. Such studies add to the effort of trying to unravel the role bat species play in the maintenance of Ebola virus.

**Uncertainty surface and sensitivity analysis**

The uncertainty surface maps showed that the suitability maps produced were robust. A variation of ± 25% introduced to the four categories of risk factors produced only a slight variation in the suitability for EBOV spillover, with a maximum average relative change in pixel value of less than 0.08. This and the small standard deviation observed indicate that the four categories of factors were a good choice for modeling EBOV spillover at the chosen local scale.

The sensitivity analysis performed to assess the effect of each risk factor on the suitability map of its corresponding category showed which factors had a more important effect in each category. In general, the most sensitive factors were those assigned higher weights, and were consistent among the three studied areas. However, there were differences on their contribution to the variation in the output map. In general, the results of the sensitivity analysis were more similar between the areas in Congo and Gabon compared to Guinée forestière, suggesting that risk factors and their interactions to promote the spillover of EBOV are not necessarily the same between these areas. For example, forest loss had a greater impact on the variation in the suitability for EBOV in Congo and Gabon than in Guinée forestière. The landscape of Guinée forestière has changed from a natural forest to a mosaic of crops and forests over the last 40 years [36,79], while Congo and Gabon still conserve large areas of less disturbed forests [37]. It is possible that the loss of forest cover is less important for EBOV spillover in already fragmented areas if for instance, bat communities have already adapted to anthropogenic landscapes through the selection of anthropophilic species and counter-selection of species more negatively impacted by human activities. Human population density also contributed more to the variation in Guinée forestière than in Congo and Gabon. Guinée forestière was more densely populated (0.36 inhabitants/km$^2$, with a maximum density of 48 inhabitants/km$^2$ in 2013), compared to Congo (0.02 inhabitants/km$^2$, max density of 0.21 in 2001) and Gabon (0.01 inhabitants/km$^2$, max density of 0.15 in 2001). A higher population density may lead to a higher probability of human-wildlife encounters in fragmented areas [80,81].

Bushmeat hunting areas and bushmeat trade contributed more to the output variation in Guinée forestière than in the Congo and Gabon study areas. On the one hand, this may result from the observation that far more villages were mapped in Guinea across the entire study area, compared to fewer villages in Congo and Gabon (see S1 Text). On the other hand, due to a lack of accurate data on bushmeat hunting areas in Guinée forestière, we assumed that forest reserves were the areas used for hunting, and reclassified them as 0 or 1. Areas around classified forests, national parks and other wildlife reserves are among those used for hunting in Guinea [67]. In contrast, although still imprecise, hunting areas in Congo and Gabon were based on estimated hunting pressure [82] reclassified from 0 to 4. Consequently, an introduced variation in weight had a more substantial effect on hunting areas in Guinée forestière than in the other study areas.

Finally, bats, particularly fruit bats, were a more sensitive parameter than primates and duikers, particularly in Congo. This is explained by the presence of five species of fruit bats in all of the mapped area in Congo, including the three species with the highest weights (Table A in S2 Text), compared to only four and two species distributed in the mapped areas in Gabon and Guinea, respectively. In contrast, the difference in the number of species in the other groups and their relative weights were less marked among the three study areas.

**Limitations of the study**

The risk maps of EBOV spillover and *Ebolavirus* maintenance in bats produced in this study, as well as their validation and the sensitivity analysis, support the use of GIS-MCE for mapping the risk of spillover of a complex zoonosis such as Ebola virus disease. Nevertheless, some limitations to our study need to be mentioned.

First, it is possible that factors that may be important for EBOV spillover were not taken into account. For example, elevation has been associated with the risk of *Ebolavirus* spillover [16,18,32], although it is not clear what role it would play in virus transmission. In addition, we grouped and assessed similar factors, such as loss of forest cover and forest fragmentation, into one. However, their association with EBOV spillover might differ. For instance, patches of forest remaining in a fragmented forest are potentially more important for contact between maintenance hosts, intermediate hosts and humans than deforested areas. Similarly, mean temperature or temperature range might not be associated with the spillover of EBOV in the same way if they are estimated on an annual or a daily basis [16,18,23,32,39]. Annual estimates are probably related to the phenology of processes such as fruiting of plants or reproductive cycles of mammals, while daily estimates are more likely representative of daily activity patterns of reservoir, intermediate and target hosts. Our choice of risk factors and the assigned weights were based on information extracted from a literature review; consultations with locals and experts on the different study areas would probably result in a more accurate classification of the risk factors.

Second, the results of GIS-MCE are to a certain extent subjective due to the amount and quality of information available [30]. The available data layers used in our model differed in their spatial resolution, from 30 m to 5 km (Table B in S1 Text), which can affect the precision of the resulting maps [83], and some of the data that we used were coarse representations of their corresponding variables. For example, the data for species distributions assumed an equal presence throughout the distribution area, but in fact, the probability of a species being present in a certain area varies and depends on several factors [84]. Thus, species distributions based on occurrence data would be more appropriate. Similarly, we did not have access to precise data on bushmeat hunting areas and bushmeat trade. Activities related to bushmeat consumption represent an important risk factor for *Ebolavirus* spillover [16], and different sectors of the population, notably hunters and women, may be more at risk [65].

Finally, our maps represent suitable areas for EBOV spillover to humans, but most of the studies reviewed identified risk factors for all Ebola viruses pathogenic to humans, without distinguishing among the species. Ebola viruses might differ in their ecological niche [16]; generalizing that the conditions leading to a spillover are the same for all Ebola viruses thus may be wrong. The relatively small number of *Ebolavirus* spillover events reported to date makes it difficult to study the contexts in which a spillover of different Ebola viruses may occur. Nevertheless, it is important to bear in mind that the risk factors and the interactions that favor a spillover of an Ebola virus might differ among the different Ebola virus species.

## Conclusion

Ebola virus disease is a current health threat that is no longer restricted to Central Africa, as the outbreak in Western Africa of 2014 and the most recent one in January 2021 made clear. Changes in climate, land use, human population density and socio-economic factors may lead to more frequent, and in some areas, more intense, EBOV outbreaks [85]. Useful tools for disease monitoring are needed to prevent and aid preparedness for future outbreaks. Here we used a GIS-MCE approach to produce risk maps of suitable areas for EBOV spillover to human populations. Our results show that spillover events were not always located in areas of high suitability. Nevertheless, given the limited data to validate our maps and the caveats mentioned above, we believe that the spillover events in Congo, located in high suitable areas, and the results of virus circulation in bats provide some support to our model. The sensitivity analysis showed that the maps produced were robust, and that there were differences in terms of the most sensitive risk factors, particularly between the study area in Guinea and those in Congo and Gabon, indicating that eco-regional rather than continental or national approaches would inform better on the relative contribution of different factors in the ecology of EBOV.

GIS-MCE is a simple approach that allows the integration of multiple criteria to solve a spatial problem. The resulting maps depend on the quality of the data and information used. Here we used only open access data to produce our geographical risk factor layers. The suitability maps produced can be easily improved and updated as more spatial data becomes accessible and more accurate, and as our knowledge of the factors involved in the spillover of the virus to the human population advances (R code provided in S6 Text). Although significant advances have been made in trying to understand the transmission cycle of the Ebola virus, we believe there are still important gaps in our knowledge of what factors lead to a spillover to humans. Notably our search for published studies using 'ebola',' ecology' and 'spillover' as

keywords found only 22 records from the Web of Science for a period spanning 30 years. Our model could be used for research and surveillance. For instance, it could inform on where research efforts should be directed for sampling campaigns of wildlife and human population to look for evidence of previous undetected outbreaks [86]. Similarly, it could be a useful tool for risk-based surveillance, highlighting areas considered at higher risk, and thus directing human and financial resources to priority areas [87].

## Supporting information

**S1 Table. Literature review results.**

**S1 Text.** Sources of data, spatial data manipulation and estimation of variables associated with risk factors of EBOV spillover.

**S2 Text. Assessment of potential reservoir and intermediate host species.**

**S3 Text. GIS-MCE applied to produce suitability maps for *Ebolavirus* maintenance in fruit bats and insectivorous bats.**

**S4 Text.** Pairwise comparison matrices of the analytical hierarchy process (AHP) for risk factors associated with EBOV spillover.

**S5 Text.** Supplementary Results.

**S6 Text. R code.**

Available at
https://journals.plos.org/plosntds/article?id=10.1371/journal.pntd.0009683#sec025